\documentclass[aip,reprint,nofootinbib,amssymb, amsmath,cha]{revtex4-1}

\usepackage{docs}%
\usepackage{bm}%
\usepackage{graphicx}%
\expandafter\ifx\csname package@font\endcsname\relax\else
 \expandafter\expandafter
 \expandafter\usepackage
 \expandafter\expandafter
 \expandafter{\csname package@font\endcsname}%
\fi

\def\simge{\mathrel{%
       \rlap{\raise 0.511ex \hbox{$>$}}{\lower 0.511ex \hbox{$\sim$}}}}
\def\simle{\mathrel{
       \rlap{\raise 0.511ex \hbox{$<$}}{\lower 0.511ex \hbox{$\sim$}}}}
\begin{document}

\title{A magnetically shielded room with ultra low residual field and gradient}%

\author{I. Altarev} \affiliation{Physikdepartment, Technische Universit\"at M\"unchen, D-85748 Garching, Germany}%
\author{E. Babcock} \affiliation{J\"ulich Center for Neutron Science, Lichtenbergstrasse 1, D-85748 Garching  Germany }
\author{D. Beck} \affiliation{University of Illinois at Urbana-Champaign, Urbana, Il 61801, USA}
\author{M. Burghoff}\affiliation{Physikalisch-Technische Bundesanstalt Berlin, D-10587 Berlin, Germany}%
\author{S. Chesnevskaya}\affiliation{Physikdepartment, Technische Universit\"at M\"unchen, D-85748 Garching, Germany}%
\author{T. Chupp}\affiliation{University of Michigan, Ann Arbor, MI 48109, USA}%
\author{S. Degenkolb}\affiliation{University of Michigan, Ann Arbor, MI 48109, USA}%
\author{I. Fan}\affiliation{Physikalisch-Technische Bundesanstalt Berlin, D-10587 Berlin, Germany}%
\author{P. Fierlinger}\affiliation{Physikdepartment, Technische Universit\"at M\"unchen, D-85748 Garching, Germany; Fierlinger Magnetics GmbH, D-85748 Garching}%
\email{pfierlinger@gmail.com}
\author{A. Frei}\affiliation{Forschungneutronenquelle Heinz Meier-Leibnitz, D-85748 Garching, Germany}%
\author{E. Gutsmiedl}\affiliation{Physikdepartment, Technische Universit\"at M\"unchen, D-85748 Garching, Germany}%
\author{S. Knappe-Gr\"uneberg}\affiliation{Physikalisch-Technische Bundesanstalt Berlin, D-10587 Berlin, Germany}%
\author{F. Kuchler}\affiliation{Physikdepartment, Technische Universit\"at M\"unchen, D-85748 Garching, Germany}%
\author{T. Lauer}\affiliation{Forschungneutronenquelle Heinz Meier-Leibnitz,  D-85748 Garching, Germany}%
\author{P. Link}\affiliation{Forschungneutronenquelle Heinz Meier-Leibnitz,  D-85748 Garching, Germany}%
\author{T. Lins}\affiliation{Physikdepartment, Technische Universit\"at M\"unchen, D-85748 Garching, Germany}%
\author{M. Marino}\affiliation{Physikdepartment, Technische Universit\"at M\"unchen, D-85748 Garching, Germany}%
\author{J. McAndrew}\affiliation{Physikdepartment, Technische Universit\"at M\"unchen, D-85748 Garching, Germany}%
\author{B. Niessen}\affiliation{Physikdepartment, Technische Universit\"at M\"unchen, D-85748 Garching, Germany}%
\author{S. Paul}\affiliation{Physikdepartment, Technische Universit\"at M\"unchen, D-85748 Garching, Germany}%
\author{G. Petzoldt}\affiliation{Physikdepartment, Technische Universit\"at M\"unchen, D-85748 Garching, Germany}%
\author{U. Schl\"apfer}\affiliation{IMEDCO AG, CH-4614 H\"agendorf, Switzerland}%
\author{A. Schnabel}\affiliation{Physikalisch-Technische Bundesanstalt Berlin, D-10587 Berlin, Germany}%
\author{S. Sharma} \affiliation{University of Illinois at Urbana-Champaign, Urbana, Il 61801, USA}%
\author{J. Singh}\affiliation{Physikdepartment, Technische Universit\"at M\"unchen, D-85748 Garching, Germany}%
\author{R. Stoepler}\affiliation{Physikdepartment, Technische Universit\"at M\"unchen, D-85748 Garching, Germany}%
\author{S. Stuiber}\affiliation{Physikdepartment, Technische Universit\"at M\"unchen, D-85748 Garching, Germany}%
\author{M. Sturm}\affiliation{Physikdepartment, Technische Universit\"at M\"unchen, D-85748 Garching, Germany}%
\author{B. Taubenheim}\affiliation{Physikdepartment, Technische Universit\"at M\"unchen, D-85748 Garching, Germany}%
\author{L. Trahms}\affiliation{Physikalisch-Technische Bundesanstalt Berlin, D-10587 Berlin, Germany}%
\author{J. Voigt}\affiliation{Physikalisch-Technische Bundesanstalt Berlin, D-10587 Berlin, Germany}%
\author{T. Zechlau}\affiliation{Forschungneutronenquelle Heinz Meier-Leibnitz,  D-85748 Garching, Germany}%

\date{\today}

\maketitle

\begin{quotation}
A versatile and portable magnetically shielded room with a field of (700$\pm$200)~pT within a central volume of 1~m$~\times~$ 1~m~$\times$~1~m and a field gradient less than 300~pT/m is described.
This performance represents more than a hundred-fold improvement of the state of the art for a two-layer magnetic shield and  provides an environment suitable for a next generation of precision experiments in fundamental physics at low energies; in particular, searches for electric dipole moments of fundamental systems and tests of Lorentz-invariance based on spin-precession experiments.
Studies of the residual fields and their sources enable improved design of future ultra-low gradient environments and experimental apparatus.
This has implications for developments of magnetometry beyond the femto-Tesla scale in, for example biomagnetism, geosciences and security applications and in general low-field NMR measurements.
\end{quotation}

\section{\label{sec1:level1}INTRODUCTION}
Magnetically shielded rooms (MSRs) provide low field ($<10^{-9}$ T) and low gradient ($<10^{-9}$ T/m) environments with strong damping of electromagnetic distortions over a wide range of frequencies.
Applications include magnetic resonance imaging at very low fields~\cite{17}, measurements of biomagnetic fields of the brain at near-DC frequencies~\cite{cohen, sander}, investigations with magnetic nano-particles~\cite{alexiou} and security applications~\cite{espy}.
Magnetic shielding is also crucial for experiments in fundamental physics, prominently in searches for  electric dipole moments of fundamental systems (EDMs)~\cite{7,8,9,ROSENBERRY}.
A finite EDM would be a manifestation of a new source of time reversal violation and should provide critical information to understand the asymmetry of matter and anti-matter in the universe and to test physics beyond the Standard Model of Particle Physics~\cite{1,2,3}, which is complementary to high-energy collider ({\it i.e.} LHC) experiments.
EDMs are usually measured using the spin-precession of one or two species of nucleons, leptons or atoms, reaching resolutions in the nHz range.
Using cryogenic SQUID-based magnetometers with sub-fT resolution \cite{17} and optical magnetometers\cite{18}  in low field and low field-gradient environments, violation of Lorentz invariance~ \cite{4,10,5,6} or spin-couplings that arise due to exotic physics can be investigated with unprecedented sensitivities.
%
%

The stability and homogeneity of the magnetic field is crucial for spin-precession measurements due to the first order coupling to spin. Effective stability of the magnetic field can be achieved with the use of a co-magnetometer, an auxiliary species often in the same volume that is less sensitive to the new physics couplings, but is comparably sensitive to the magnetic-moment coupling; however the homogeneity generally affects the spin-coherence time ($T_2^*$) or observation time for a measurement in first or second order. Generally the longest spin-coherence time requires the smallest gradient~\cite{14}.

The state-of-the art is the Berlin Magnetically Shielded Room 2 (BMSR-2)\cite{19} at the Physikalisch Technische Bundesanstalt (PTB) Berlin.
With seven layers of $\mu$-metal and a mass of about 24 tons, BMSR-2 has a passive shielding factor (SF) of $\sim$~75000 for external magnetic field variations at frequencies less than 0.01~Hz. 
With additional active shielding, external variations are attenuated by a factor of more than one million. 
In this paper,  we describe a transportable MSR developed as part of the magnetic shielding for an experiment to measure the EDM of the neutron\cite{16} at the FRM-II reactor of the Technische Universit\"at M\"unchen (TUM) in Garching.
With just two layers of $\mu$-metal and an RF shielding layer, the transportable room has a shielding factor of $\sim$~300 for external magnetic field variations at frequencies less than 0.01~Hz. 
The residual static field and static-field gradients are nevertheless comparable.

\section{\label{sec2:level0}APPARATUS}

\subsection{\label{Design Principles} Design Principles}


To achieve the goals of small field, small gradient, accessibility and vibration isolation, we applied the following design principles to our MSR:

\begin{itemize}

\item
Physical placement of the room mechanically and electrically decouped from the enclosing building. A partly magnetic decoupling could be achieved by cutting out the steel reinforcement in the MSR and replacing it by a seperate stainless steel reinforcement.  

\item
A controlled magnetic environment within which the MSR is placed inside an active surrounding-field compensation (SFC) for both static control of e.g. the Earth's field and dynamic stabilization of fields varying slowly in time, also called active shielding.

\item
A combination of  magnetic and RF shielding using highly permeable and highly conducting materials. The large inner dimensions require assemblies of several plates with connections, with the plates as large as possible.

\item
Access to the inside is provided by a large removable door. Effects on shielding performance and magnetic gradients inside the MSR are minimized by placing the door-wall-overlaps close to the edges of the room, far away from the center.

\item Permanently installed optimized degaussing system for reproducible and complete degaussing to low residual field values.

\item
Negligible magnetic effects from all internal equipment including lighting.

\item 
Strict magnetic screening for all materials and fasteners used in constructing the room and for the experimental apparatus used within the room. In the present case, samples of all materials are screened inside BMSR-2 for magnetic contamination. Specifications are less than or equal to 3~pT at 3~cm for items closest to the actual experiment inside.
\end{itemize}

Note that any magnetizable piece within the shield, even if initially demagnetized could be magnetized by a field fluctuation, the B$_0$-field, or sparks from electromagnetic discharges.








\subsection{\label{sec2:level1} Field compensation and active shielding}

The FRM-II neutron EDM experiment is located in the Neutron Guide Hall East building.  
To accommodate the neutron-beam height, the base for the MSR is set in a 9~m $\times$ 6~m sized and 1~m deep concrete basin.  
The basin has a double-wall and is electrically and mechanically isolated from the building's floor. 
The concrete is reinforced with non-magnetic stainles-steel rods.  

Surrounding the MSR, resting on the floor of the basin, is a field-cage system consisting of 24 rectangular coils with dimensions X=6~m $\times$ Y=9~m $\times$ Z=6~m~\cite{4a}.
This coil system is used to apply linear and second order gradients to reduce the magnitude and inhomogeneity of the magnetic field in the vicinity of the MSR.
The same coils can be also used to apply fields and gradients to test the magnetic properties of the MSR.

\begin{figure}[tb]
\includegraphics[scale = 0.31]{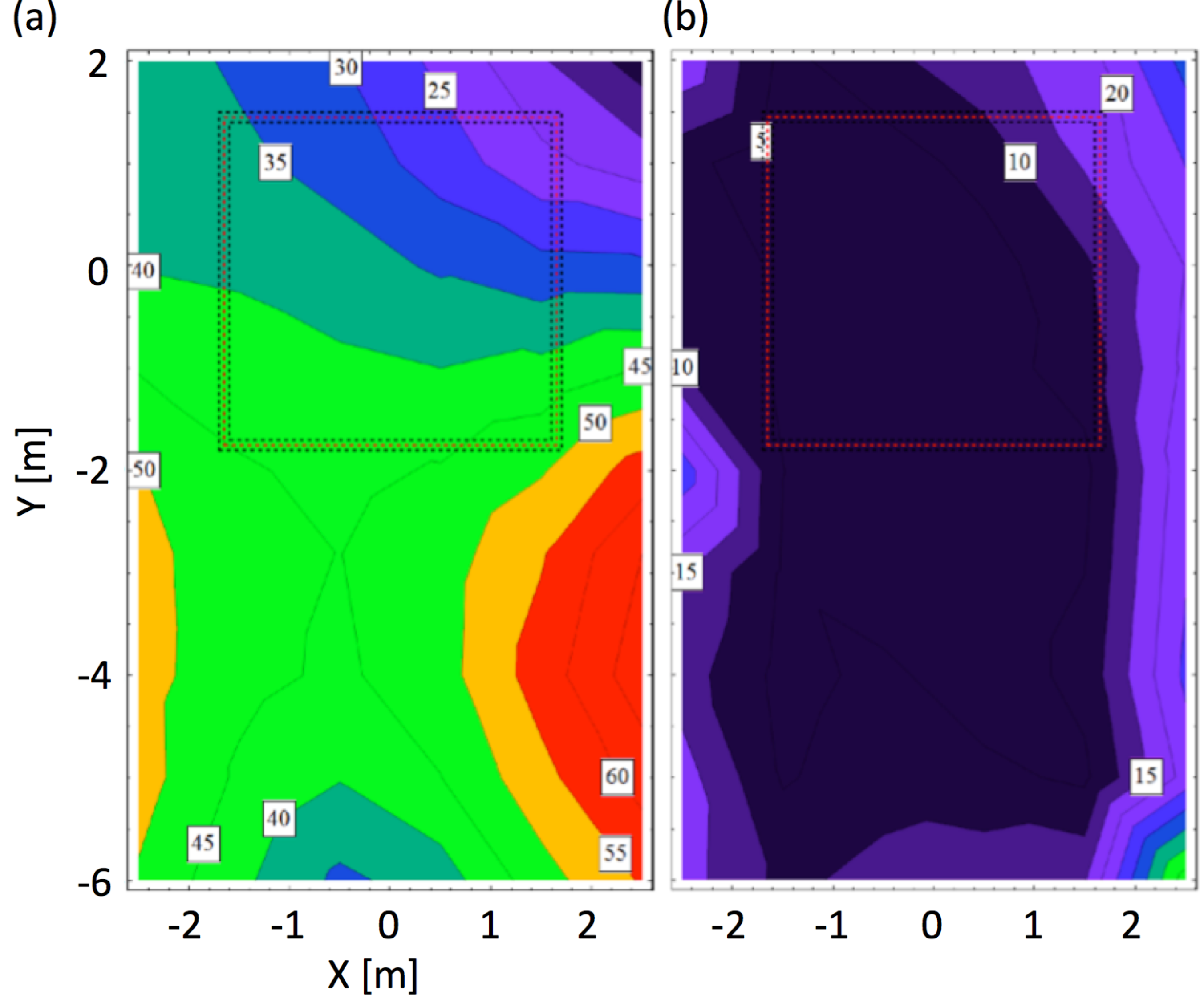}
\caption{\label{fig:cage}  A horizontal cut through the (a) ambient field, and (b) field-cage compensated magnetic field at a height $Z=1.5$~m, approximately the center of the MSR. The dotted lines show the position of the walls of the MSR. The contour labels indicate the magnetic field in $\mu$T.
}
\end{figure}

Low-frequency, time-dependent distortions, like cars passing nearby the experimental hall, the Munich subway line to Garching, and elevators in the building, can be actively compensated at frequencies up to 10~Hz using feedback to the 24 coils calculated with data from an array of 60 3-axis fluxgate magnetometers~\cite{bartington}.  
The fluxgate magnetometers are located within the volume of the field-cage, but outside the MSR, and are arranged in a pattern to optimize the off-center MSR location within the coil system (see Fig.~\ref{fig:cage} and Sec.~\ref{sec2:level2})~\cite{4a}. 
The effect of the MSR on the surrounding field is taken into account by the compensation algorithm.
Without the MSR in place, the ambient magnetic field at the position of the experiment  with and without the field compensation is shown in Fig.~\ref{fig:cage}.


\subsection{\label{sec2:level2}The MSR construction}

\begin{figure}[tb]
\includegraphics[scale = 0.36]{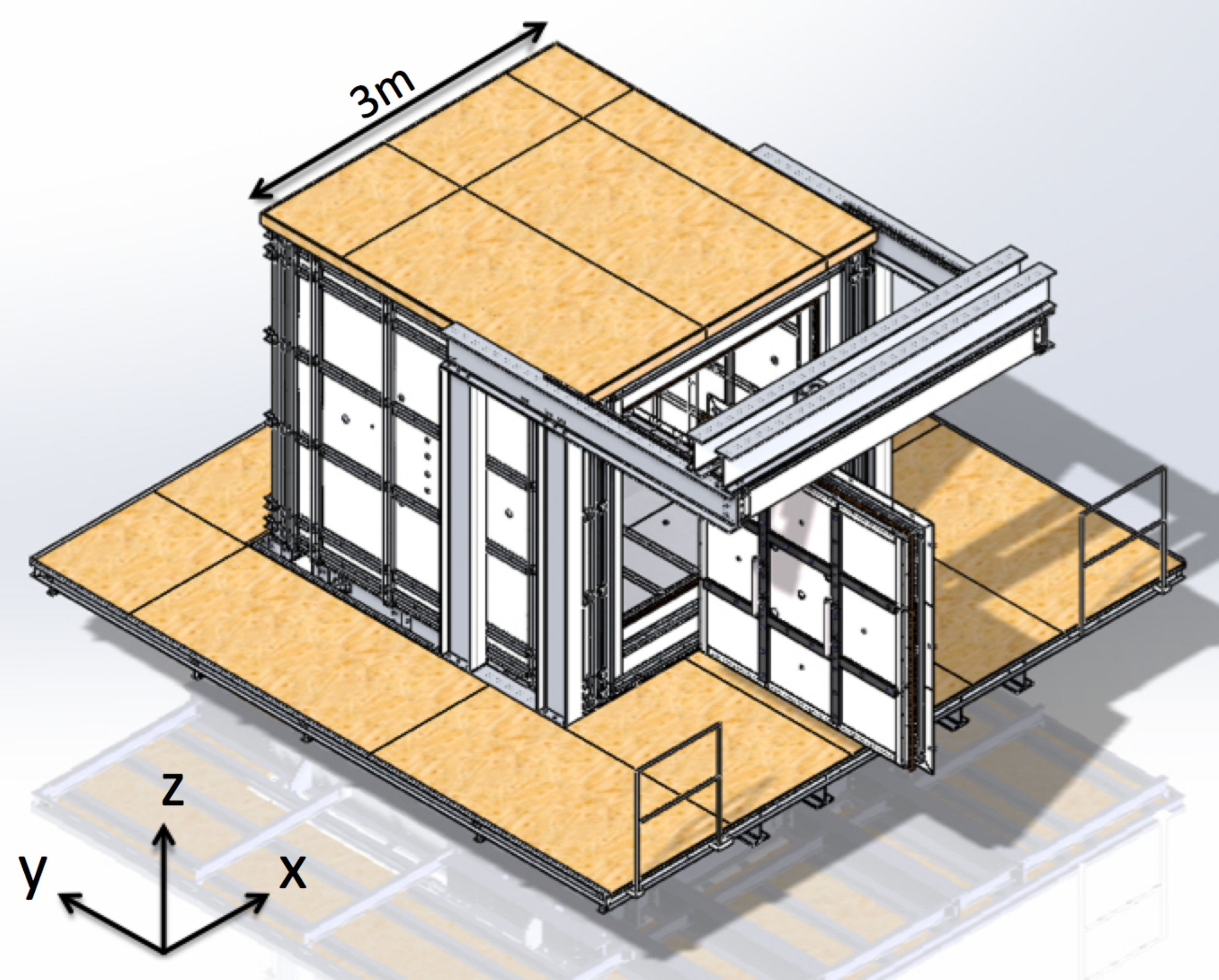}
\caption{\label{fig:msr} 3D model of the MSR. The MSR is placed on an aluminum table, the floor level in the room and the surrounding platform is 1~m above ground.
At the front side, the $2\times2$~m$^2$ access door is shown opened, but not fully retracted. The MSR is purposely placed close to the side of the coil system so that (a) the distance between the experimental area and the neutron source is minimized and (b) the door of the MSR can be opened freely.
}
\end{figure}

A drawing of the MSR and the surrounding wooden access floor is shown in Fig.~\ref{fig:msr}.  
Both the access floor and the room floor are at the ground level of the experimental hall, 1~m above the bottom of the concrete basin.
The MSR provides an inner space of dimensions $2.50\hbox{ (X)}\times2.78\hbox{ (Y)}\times2.30\hbox{ (Z)}$~m$^3$ (see Fig.~\ref{fig:msr}) and weighs about 9200~kg.

\begin{figure}[tb]
\includegraphics[scale = 0.41]{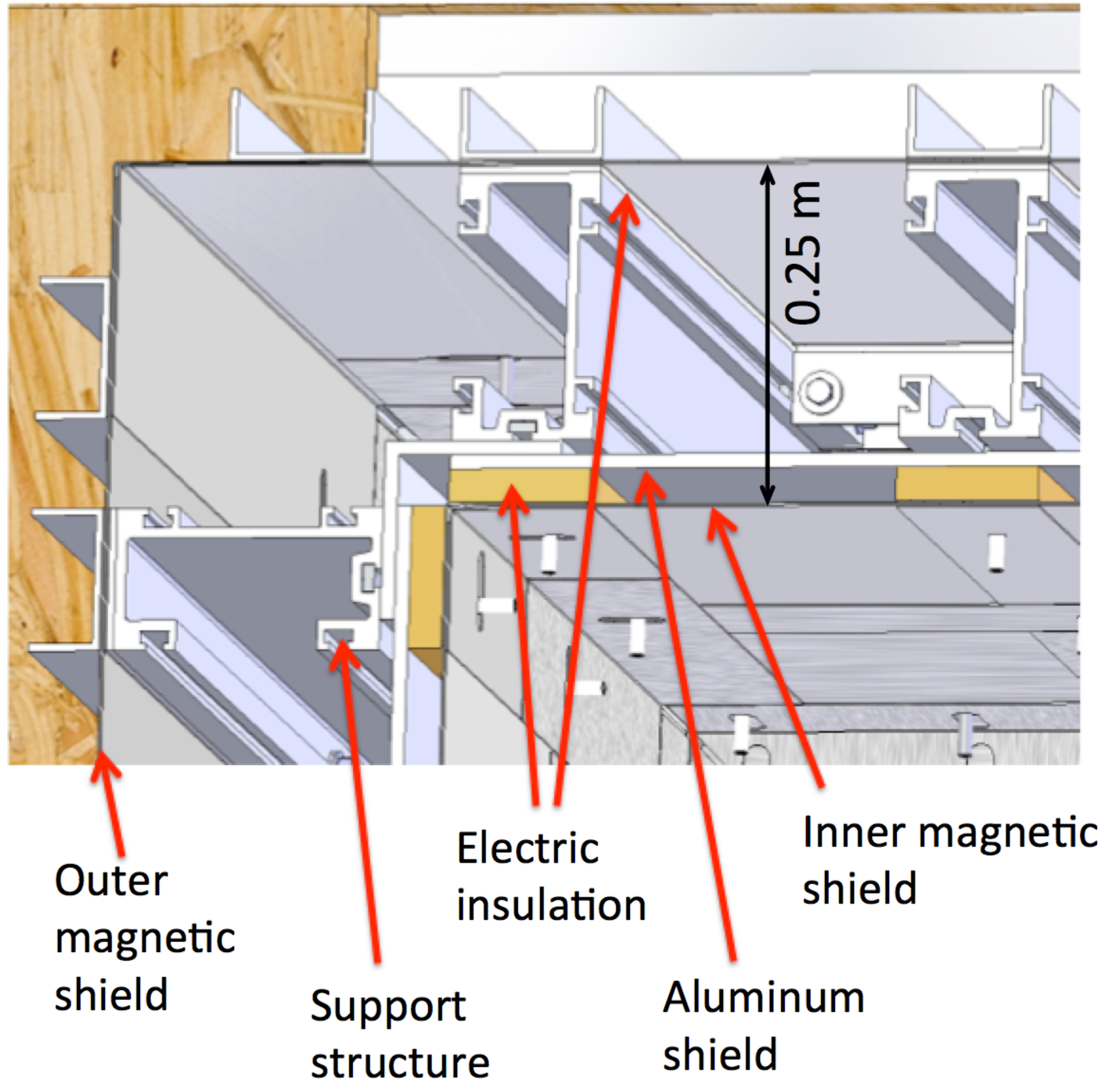}
\caption{\label{fig:wallsegment}  Typical section of the walls of the MSR. Note that there is also electric isolation (not shown) between the outer layer, the aluminum layer and the inner layer.
}
\end{figure}

The walls of the MSR consist of three functional layers shown in Fig.~\ref{fig:wallsegment}: inner and outer passive magnetic shielding layers for low frequency electromagnetic fields and an aluminum mid-layer for  eddy-current magnetic field damping and RF suppression.  
Each passive shielding layer contains two 1~mm thick $\mu$-metal sheets, fabricated from Krupp Magnifer\cite{KruppMagnifer}, a high permeability alloy consisting of mainly nickel and iron\footnote{We will continue using the commonly used term ``$\mu$-metal'' for this particular brand throughout this paper}. 
The two sheets of each layer are mounted in a crossed arrangement with high symmetry to avoid directional dependencies of the gaps in the assembly.
Additional strips of $\mu$-metal of a width 70~mm are used to cover gaps on each side. 
Holes for mounting screws are placed in regular 25~cm hole patterns of maximum 75~cm separation in all  $\mu$-metal sheets.
Stacking more sheets of $\mu$-metal in one layer helps to match the magnetic field used for degaussing to penetrate the material. 
Possible gaps between the sheets in the layer do not cause any measurable changes in the performance.
The inside and outside shielding layers are separated by approximately 0.25~m as shown in Fig.~\ref{fig:wallsegment}.
All layers are electrically isolated from each other using plastic sleeves around all mounting screws and plastic layers between the support structure and each shield layer.  
This allows each functional layer to be connected separately to a single  ground point.

\begin{figure}[tb]
\includegraphics[scale = 0.40]{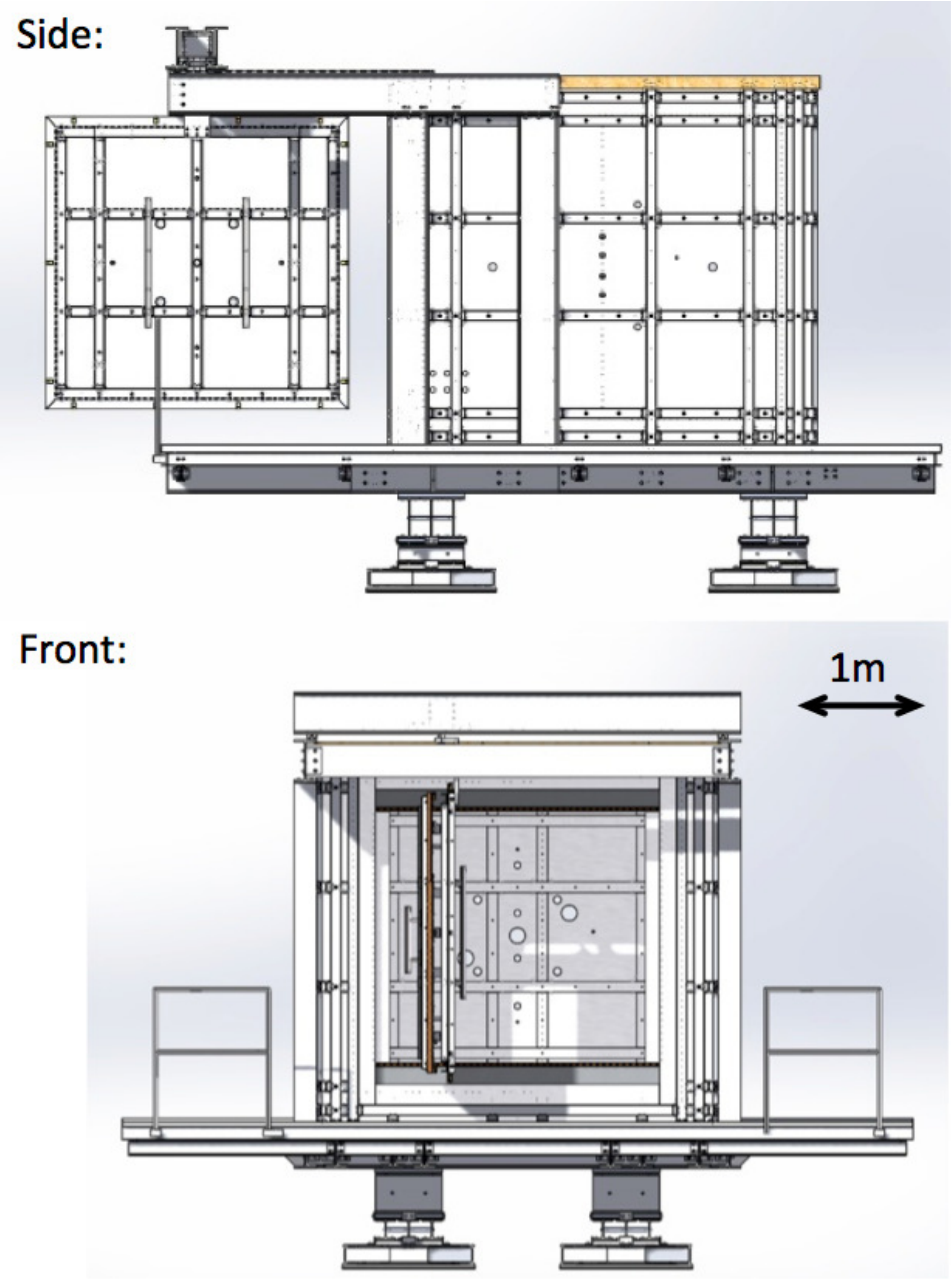}
\caption{\label{fig:side} Front and side views of the MSR with the door open. Arrays of access holes with diameters ranging from $d$~$=$~40~mm to $d$~$=$~50~mm and are placed on all sides, as well as on the top and bottom. There are three 130~mm diameter holes in the back wall (looking into the room) providing access for two neutron beam guides and a high voltage connection for the neutron EDM experiment. An aluminum structure supports the door from above, allowing it both to move in the X and Y directions and to rotate freely and to move completely out of the way for the insertion of large components.
}
\end{figure}

The  RF shielding layer is 8~mm thick aluminum and is tin-coated to provide better electrical contact at overlaps.
Feed-throughs are made from aluminum tubing 0.25~m long welded into the RF shield with holes in the $\mu$-metal sufficiently large to prevent electrical contact between layers.
There are 74 RF-tight circular holes penetrating the  walls of the MSR.  
They range in size from 20 to 130~mm diameter and may be used for cables, ventilation, vacuum lines, lasers, magnetic field sensors,  spin-polarized gases, etc..
Figure \ref{fig:side} shows the positions of these access holes in front and side views of the MSR.  
The holes are placed in symmetric patterns on opposite walls.
The penetrations have minor effects on the shielding factor and field distribution inside the room, as discussed in Sec.~\ref{sec3:level0}.
The entire MSR, including the door (see Sec.~\ref{sec2:interior}), can be transported without any additional support structure by truck. It was moved from the manufacturer~\cite{imedco} in Switzerland to Munich, and then again across the TUM campus, with no measurable increase in the residual field quality inside the MSR nor the SF.

\begin{figure}[tb]
\includegraphics[scale = 0.44]{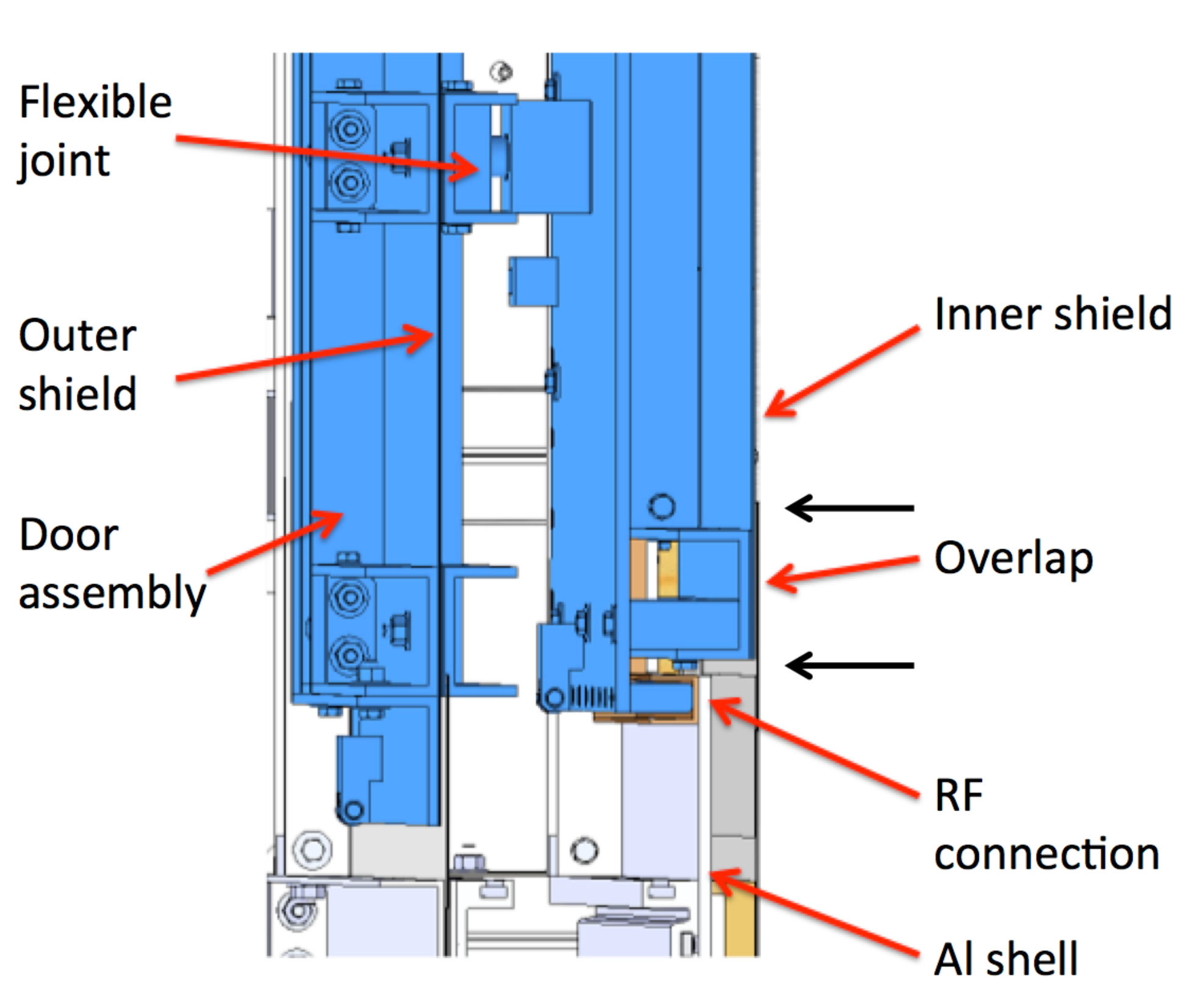}
\caption{\label{fig:door} Detail of the door  (blue) and its connection with the front wall of the room.
The overlap of the shielding material of door and wall is 65~mm for both the inner and outer shell. The aluminum shell is connected to the door via flexible copper contacts. The black arrows mark the region where the shielding material thickness is doubled due to the overlap of inner wall and door, as discussed in detail in Sec.~\ref{sec3:level0}.
}
\end{figure}

%
%


\subsection{\label{sec2:interior}The MSR interior}
The MSR is designed to accommodate a large experimental apparatus of approximate size $1.9\hbox{ (X)}\times2.8\hbox{ (Y)}\times1.9\hbox{ (Z)}$~m$^3$. 
Access is provided through a door $1.92 \hbox{ m wide} \times 2.00$~m high. 
The door is placed in the center of the front wall of the room. 
It is actuated manually with pneumatic clamps that provide a constant pressure between the shielding material of the door and that of the room.
Two sets of these clamps press the inner door shield to the inner room shield, and independently the outer door shield to the outer room shield.
A horizontal cut through the closed door is shown in Fig.~\ref{fig:door}.
The door is mounted on a rail system at a single point at the door's top center.  
The rail system is mounted on the roof of the MSR, and the door can be retracted and rotated to leave the opening completely clear.
This operation can be performed by a single person. 
An emergency mechanism allows the door to be opened from the inside, so that personnel  can work inside the room with the door closed.
Good magnetic contact at the door is critical to obtain a large SF, particularly for frequencies below about 1~Hz.
For this purpose, the magnetic material at the door has a 65~mm overlapping region which is clamped together using pneumatic actuators.  
Another important consideration of the door is that the SF at frequencies between 1 and 100~Hz in X- and Z-directions are limited by the conductivity provided by the flexible copper door contacts.

Another rail system on the floor inside the MSR allows for the insertion of experimental apparatus with a mass up to 5500~kg.
The load is transferred to a pattern of 28 supports that penetrate through the shielding layers to the shield supports 
The shield and the apparatus are mechanically coupled so that they vibrate together, further minimizing the impact of mechancial resonances.  
The major resonance frequencies for mechanical vibrations of the  room were measured to be 8, 10, 16, 20, 22.3~Hz.

A wooden floor is installed inside the room to allow convenient access to the experiment. 
The floor is supported by the shield frame, but is separated from the inner $\mu$-metal layer.
Lights made from arrays of direct current driven light emitting diodes provide standard workspace-quality illumination inside the MSR with negligible impact on the measured residual fields (see Sec.~\ref{sec3:residualField}).

\subsection{\label{sec2:level3}Degaussing }
It is possible to magnetize $\mu$-metal due to its finite remanence. To reach the lowest possible residual field inside a MSR degaussing is necessary. It turned out, that in the past the possible residual field was limited by the used degaussing procedure. We used an optimized procedure similar to the degaussing described in Ref.~\cite{12}.
The degaussing coils, consisting of 21 turns of 2.5~mm${^2}$ copper wire, are placed around each of the 12 edges of a $\mu$-metal layer.
To decrease inductance, three of the wires in each coil are connected in parallel, forming a seven-turn coil along each edge.

\begin{figure}[tb]
\includegraphics[scale = 0.31]{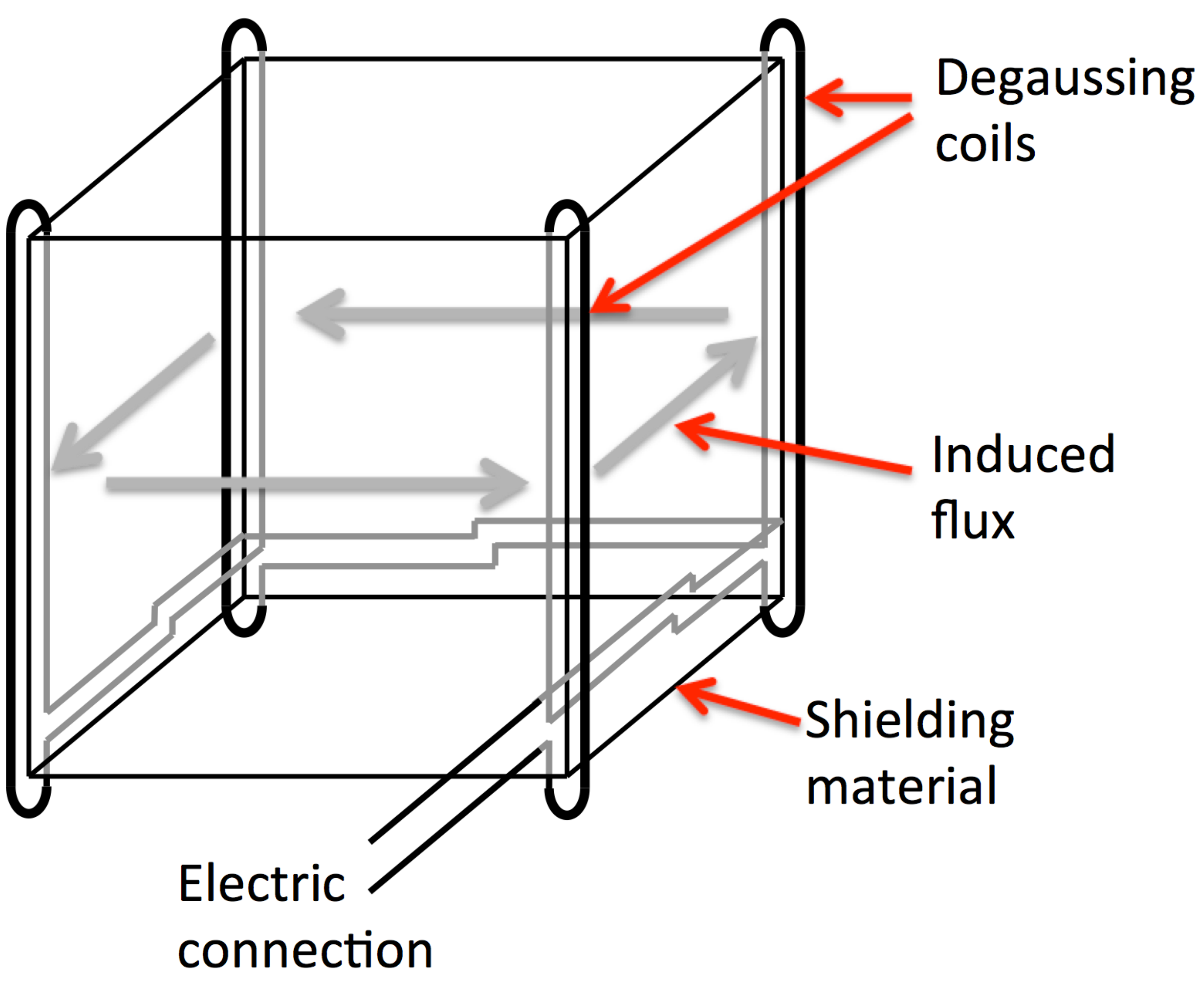}
\caption{\label{fig:degaus} Degaussing coils are placed surrounding the edges of each shell of shielding material. The figure shows a set of four coils, connected in series, for one direction in space; there is a similar arrangement for each spatial direction. The interconnecting wires are routed (and twisted) to compensate additional magnetic fields. The grey arrows denote the induced magnetic flux in the shield by the coils.
}
\end{figure}

Four of the twelve edge coils are configured in series to degauss in one spatial direction as illustrated in Fig.~\ref{fig:degaus}.
The four edge coils create a magnetic flux in a closed loop pattern inside the $\mu$-metal layer degaussing the four walls that it passes through.
To degauss the remaining two walls, another set of four edge coils around a different spatial direction has to be used after the degaussing of the first four walls.
It has been demonstrated~\cite{12} that such a degaussing procedure can lead to very low and homogeneous residual fields.

To achieve the smallest possible residual fields and gradients inside the MSR by degaussing of the MSR, several conditions must be satisfied.  
In the optimal situation, the field during degaussing at the outside of the MSR is stable, although not necessarily zero. 
Everything inside the inner magnetic shielding layer of the MSR, including fasteners and lightings, should be as non-magnetic as possible.   
The degaussing procedure then produces minimal magnetic fields in the inner magnetic shield, resulting in low residual fields and gradients within the room.  
Ultimately, in the case where the experimental apparatus is generating its own magnetic field by a time stable current through a coil inside the MSR, the degaussing should be done with the internal field switched on to get a time stable magnetic field inside the MSR.   
Degaussing is done in one direction in one shield layer at a time by applying a sinusoidal current signal with decreasing amplitude.  
In this case we use a sine function with a frequency of 10~Hz and start above saturation current which was 9~A for the inner shield. It was followed by 1000 cycles at 10~Hz with a linearly decreasing envelope.
The function is generated in software on a PC and converted to a voltage signal with a 10 kHz, 16-bit digital-to-analog converter (DAC), followed by a voltage divider to match the maximum current to the full range of the DAC.  
The signal goes through a 100~Hz low pass filter to smooth the steps of the DAC and is then converted to current by an amplifier before passing through a transformer to eliminate any d.c. offset.  
To minimize the residual field, the degaussing procedure described above is applied in turn to the magnetic flux loop (see Sec.~\ref{sec2:level3}), corresponding to each of the three spatial directions, first for the inner shell, then the outer shell and then for the inner shell a second time.  
The order of processing for the three loops of each shell is found to be unimportant for the residual field value achieved so far.  
The full non-optimized degaussing procedure is performed in less than 30~min.  
After this initial degaussing of both layers, similar residual fields were obtained by degaussing only the inner layer in $\sim$~5~min.  
For example, the residual magnetic field in the center of the chamber can be reproduced well within 0.5~nT independent of the direction or magnitude of the surrounding magnetic field of up to 100~$\mu$T or applied field gradients of up to 100~$\mu$T$/$m.

\section{\label{sec3:level0}Measurements}

\subsection{\label{sec3:level1}Shielding factors}
The performance of a magnetic shield is typically characterized by measuring the change of the field within the shield when the external field changes. For an external field provided by current applied to a test coil, the shielding factor (SF) can be defined as 

\begin{equation}
SF = \frac{\rm{test~coil~magnetic~field~without~MSR}}{\rm{test~coil~magnetic~field~with~MSR}}. \label{eq:sf}
\end{equation}
The shielding factor was measured at the center of the room for a range of frequencies and amplitudes.

\subsubsection{\label{sec3:level1a}Sudden surrounding magnetic field disturbance}

\begin{figure}[tb]
\includegraphics[scale = 0.31]{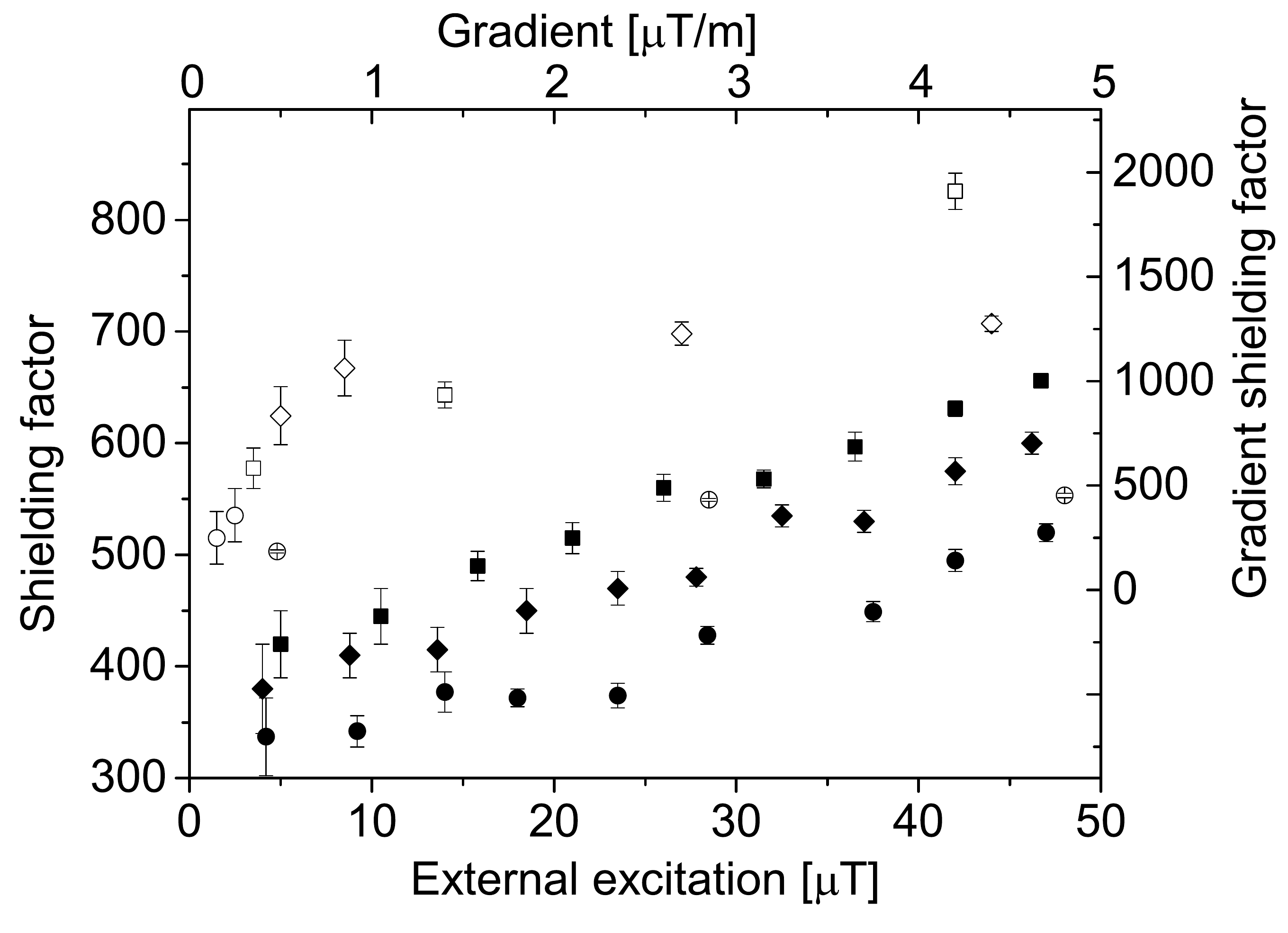}
\caption{\label{fig:fig2} Shielding factor (SF) of instantaneous steps in external homogeneous fields and gradients for different magnitudes as measured at the center of the shield. 
The filled symbols indicate the SF for homogeneous external fields: left and bottom scales (squares, X direction; diamonds, Y direction; circles, Z direction). The open symbols indicate the corresponding SF for external gradients: right and top scales. 
}
\end{figure}

When the magnetic field environment changes suddenly, such as moving of an object made of magnetic material ({\it e.g.} crane, car or door), the magnetic field inside the room will be affected.
To study this, a series of sudden changes or ``instantaneous steps'' in the external field and gradient were applied using the coils of the external compensation system (see Sec.~II-A).  
The field or gradient at the center of the shield was measured with a Mag03-70 low-noise Bartington fluxgate magnetometer.~\cite{bartington}  The SF for instantaneous uniform field changes  is greater than 300 for few-$\mu$T fields and increases with amplitude as shown in Fig~\ref{fig:fig2}.  
The smaller SF for external fields in the Z direction is due to the placement of the external coils, which are vertically offset by 0.5~m relative to the MSR; 
however, the Z-direction SF is reduced by less than a factor of two when the external field is changed by approximately 50 $\mu$T (the approximate magnitude of the earth's magnetic field).
Note that although the direction of an applied distortion is not necessarily conserved inside an MSR, here $\vert B\vert$ is used for the calculation of SF.
After a sudden external change, the internal field responds with an observed timescale of 0.6 - 1.5~s.
The increase of the SF with larger magnitude of excitation is expected from the behavior of $\mu$ as function of magnetization. Active procedures like shaking (see e.g. Ref.~\cite{budker}) of the shield will significantly increase the SF of the MSR at low frequencies.

\subsubsection{\label{sec3:level1b}Sinusoidal external magnetic field disturbance}

\begin{figure}[tb]
\includegraphics[scale = 0.28]{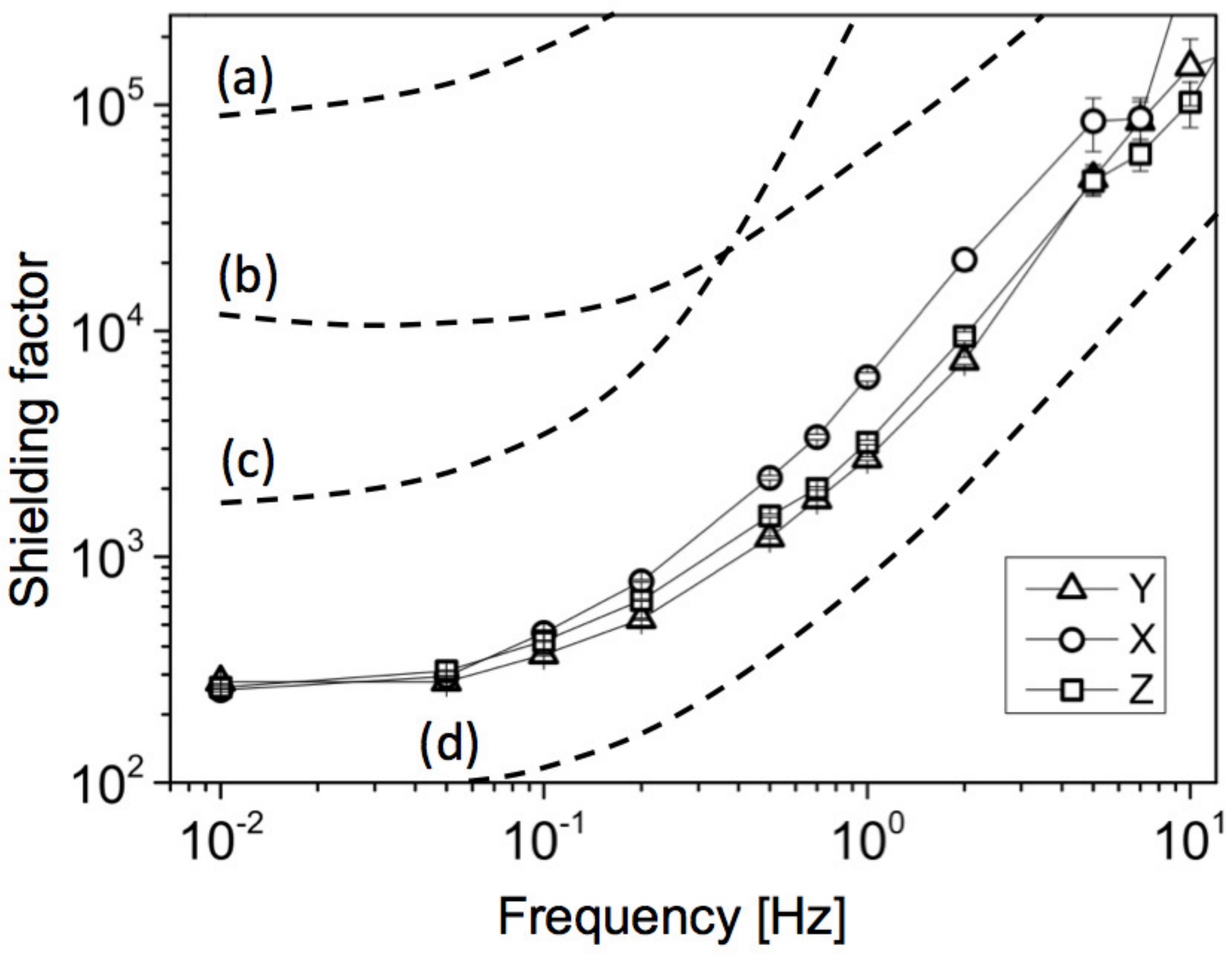}
\caption{\label{fig:shield1} The shielding factor (SF) as a function of frequency. An external, rectangular coil 
pair configuration was used to apply sinusoidal fields with a peak-to-peak amplitude of 1~$\mu$T as measured at the location of the center of the MSR prior to its installation. For frequencies above 5~Hz, the measurement uncertainties are dominated by the noise of the fluxgate probe.  The solid lines between the measured points are to guide the eye; the dashed lines show the SFs of other MSRs as follows: (a) 7-layer BMSR-2
(b) 6-layer BMSR,
(c) Boston 3-layer MSR and
(d) Ak3b MSR from Vacuumschmelze with 2 $\mu$-metal layers and one RF shield.}
\end{figure}

Sinusoidal external magnetic fields were generated by a set of rectangular coil pairs of size 4~m $\times$~5~m, with 4~m spacing, which also were used during commissioning at the factory. 
The SF as a function of frequency for low frequencies is shown in Fig.~\ref{fig:shield1}.
For frequencies below 0.01~Hz, the SF approaches 300.

Our results are compared to the performance of the following MSRs in Fig.~\ref{fig:shield1}:
(a) BMSR-2~\cite{19} with 24~tonnes of $\mu$-metal~\cite{Mumetal} in seven layers (with thicknesses, starting from the inner layer, of 4~mm, 7~mm, 6~mm, 3~mm, 3~mm, 2~mm, 2~mm),
(b) BMSR with 10 tons of $\mu$-metal in six layers;~\cite{19},
(c) the MSR in Boston~\cite{20} with three $\mu$-metal shells (inner layer 4~mm, middle and outer layer each 3~mm thick), and
(d) a typical Ak3b MSR from Vacuumschmelze \cite{ak3b}, with two layers of 4~$\times$ 0.75~mm $\mu$-metal (overall about 50\% more material than the Garching MSR presented here). 
The performance of an installation of a modified Ak3b MSR with similar dimensions and optimized degaussing is also discussed by Ref.\cite{12}. 
Note that the different shielding factors presented here are hard to compare precisely because they depend on the strength of the external field, the positioning of the probe in the room, and the geometry and distance of the test coils away from the outer magnetic shielding layer.

\begin{figure}[tb]
\includegraphics[scale = 0.32]{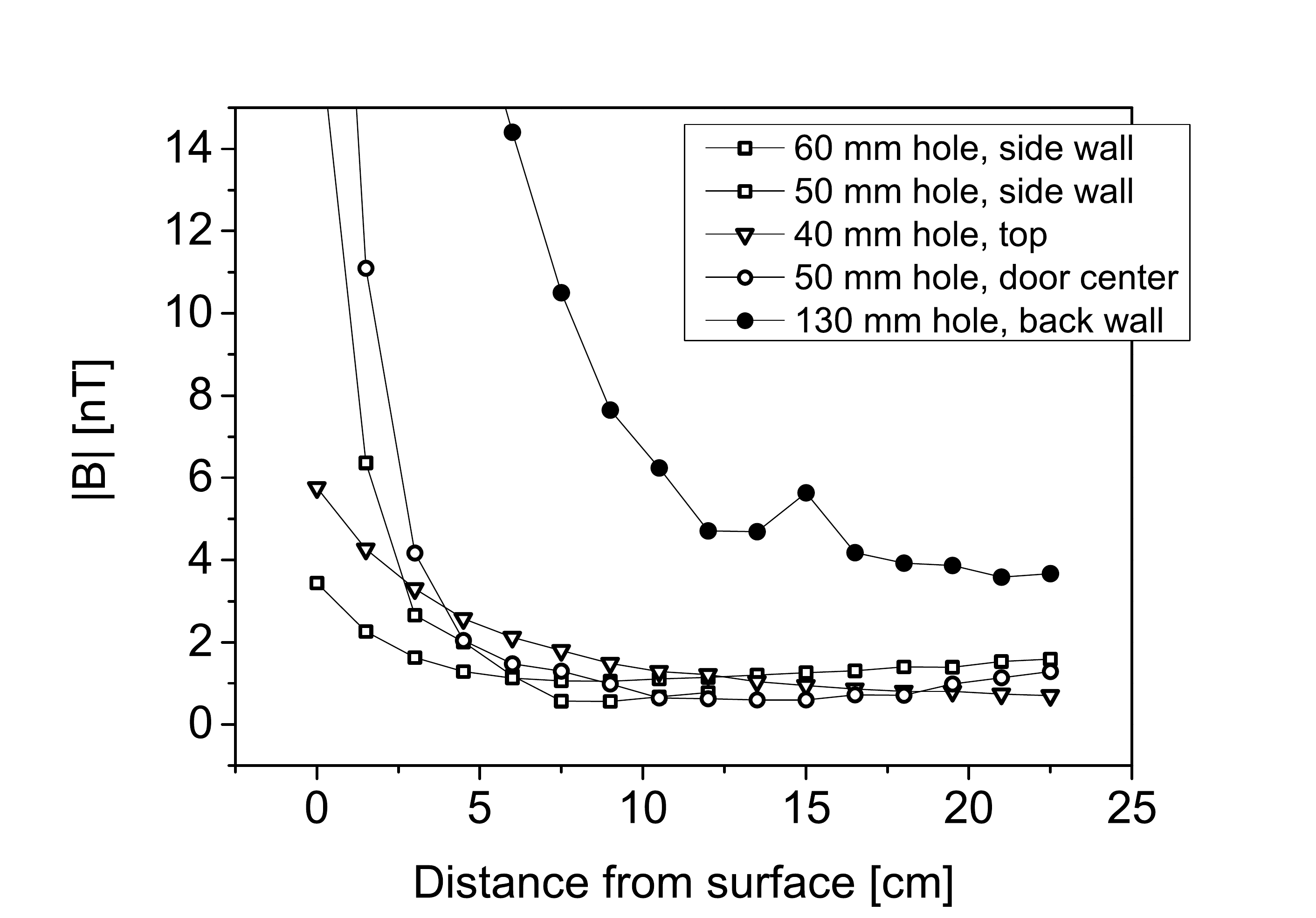}
\caption{\label{fig:fielddistance} Magnitude of the interior magnetic field as function of distance from the wall. The curves show the magnitude of the residual field measured along a line normal to the shielding material surface and centered on the corresponding feature as indicated.  The values are reproducible at the level of about 1~nT (limited by the accuracy of the fluxgate sensor) after repeated degaussing over a period of several weeks. Note that  distance = 0 corresponds to the Fluxgate-probe placed in the center of a hole in the surface of the $\mu$-metal, with the actual probe placed 10~mm away inside the housing. The map is corrected for the positions of the respective probes inside the fluxgate.
}
\end{figure}

\begin{figure}[tb]
\includegraphics[scale = 0.45]{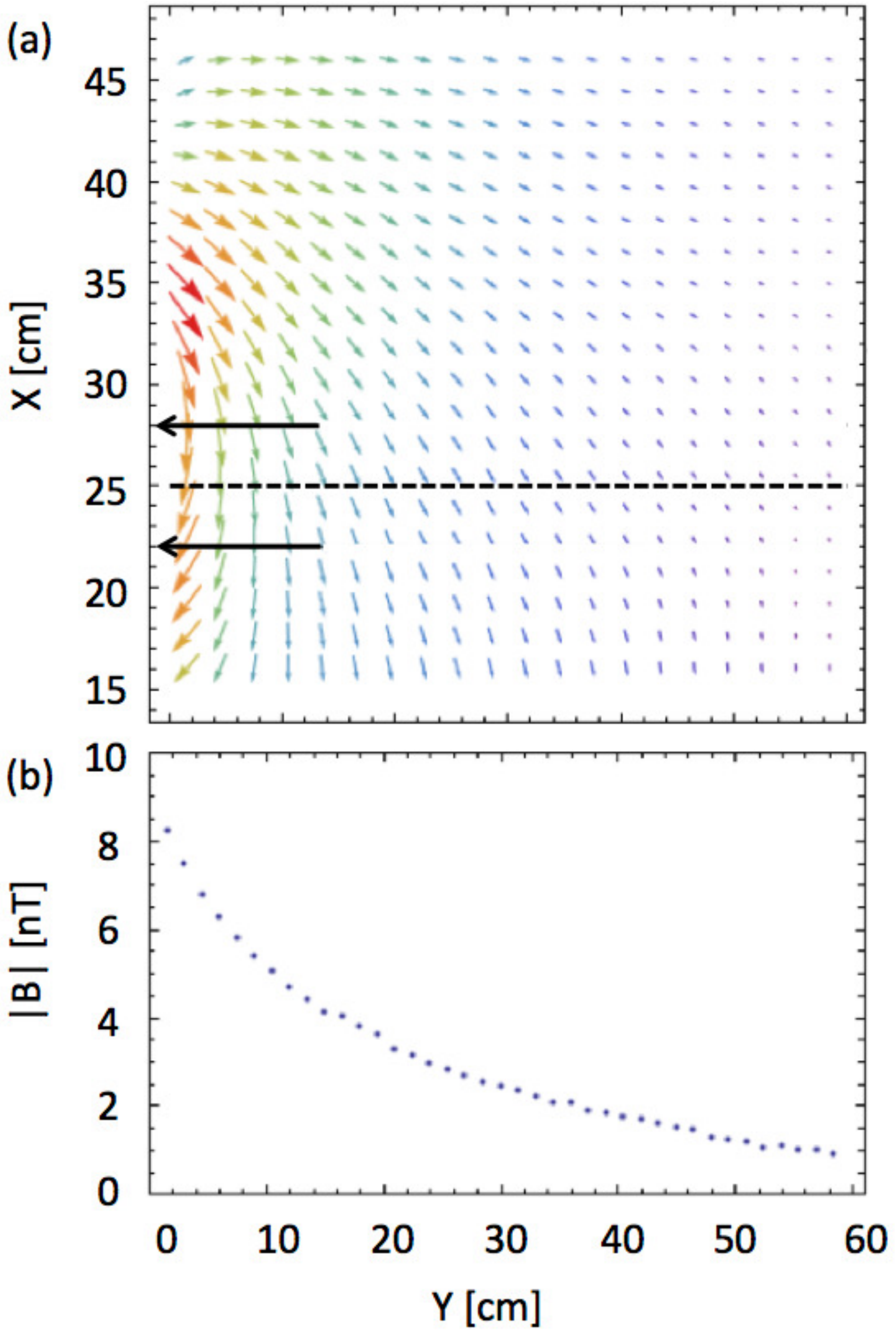}
\caption{\label{fig:fielddistance2} (a) Horizontal field on a horizontal plane in the region of the door. The arrows indicate the region of the overlap, where the inner door is pressed on the inner shield from the outside as sketched in 
Fig.~\ref{fig:door}. The dotted line indicates the path along which $\lvert B \rvert$ is plotted in (b). 
X = 0, Y = 0 corresponds to the front, right coner of the inner $\mu$-metal layer.
}
\end{figure}

\subsection{\label{sec3:residualField}Residual field}

\begin{figure}[t]
\includegraphics[width=1\linewidth]{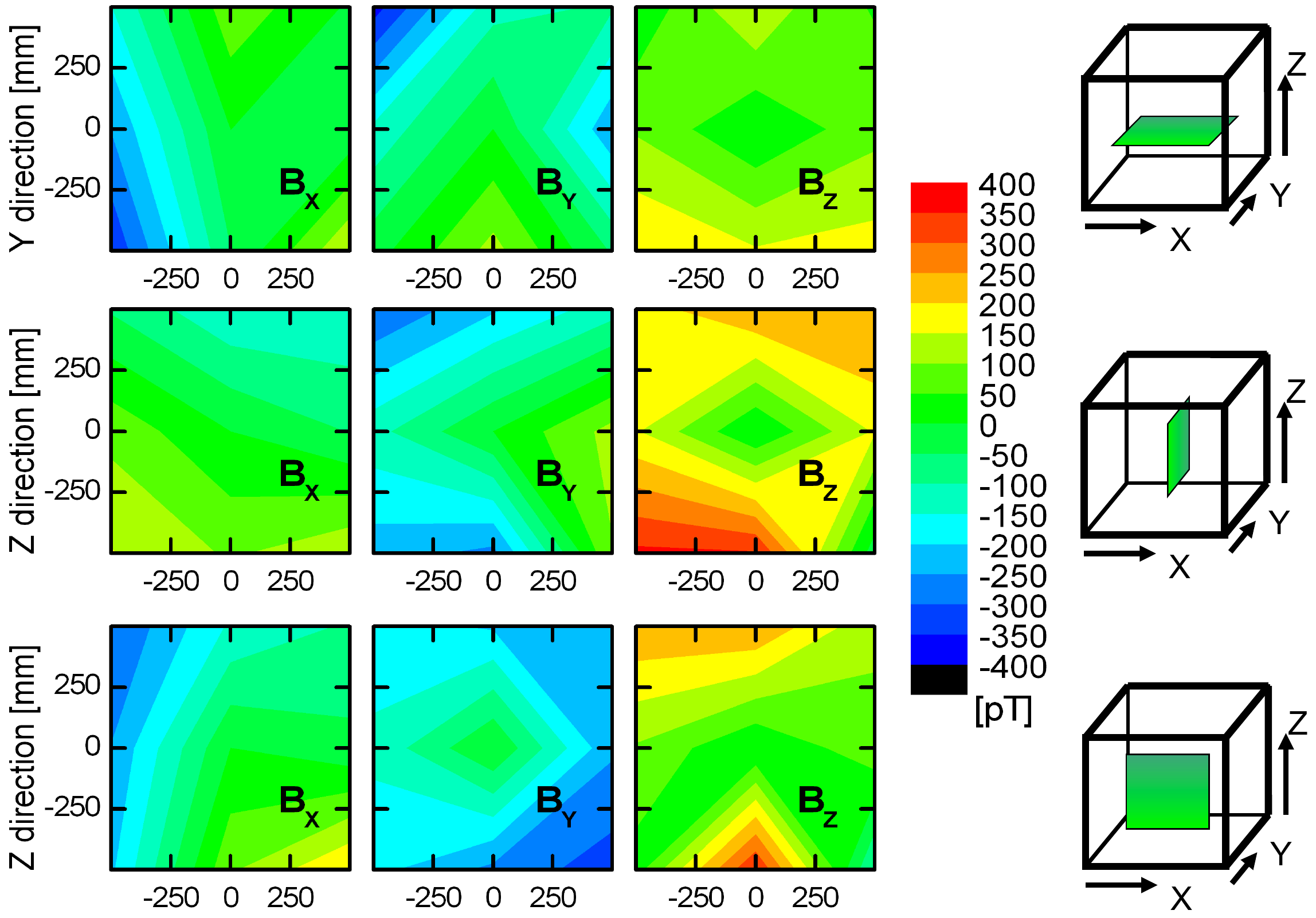}
\caption{\label{fig:field3d} Plots of the residual field homogeneity in a 3x3x3 points grid with a 0.5 m spacing where the field amplitude is normalized to the central field. The cross-sectional views are visualized on the horizontal and two vertical cuts passing through the center point. The step size of the magnetic field strength between each contour line is 50 pT. }
\end{figure}

The residual field inside the room was measured with an array of  6 LTc SQUID magnetometers~\cite{drung2003}, each mounted on one of the six faces of the cubic holder. The SQUID dewar was moved to several positions in the room. Figure~\ref{fig:field3d} shows a map of the residual field x, y and z components. 
In each measurement plane, we measure the center and center front point twice, once at the beginning and once at the end, to check the reproducibility.
Within the investigated volume, the $\Delta B_i/\Delta j$ is conservatively estimated to be less than 700~pT/m, where $i,j=x,y,z$.
At the center, the resolution limit of the magnetometer allowed us to set an upper limit on the gradient of $<$~300 pT/m.
The degaussing and the measurements were undertaken at night without the dynamic field-compensation when variations of the external field were less than 100~nT.

During these initial commissioning measurements reported here, there was a contact problem of the inner $\mu$-metal shield at the lower edge of the door as seen from Fig.~\ref{fig:field3d}.
Due to the dominance of this systematic error, the drift of the SQUID magnetometer of 150~pT during the 2 hour duration of this measurement is not corrected.
Even with this magnetic-field leakage, our measured residual field was (700$\pm$200) pT within the measured 1 m$^3$ volume and about half in the upper half of the volume.
This is the smallest value that has ever been achieved over such a large volume for a two-layer MSR (for a comparison, see Fig. 1 in Ref.~\cite{12}) . 
Due to the small SF at low frequencies, these results can only be achieved at night when the low frequency disturbance is small.
Note that the field is dominated by nonlinearities caused by the door, visible near the boundary of the volume placed exactly in the center of the room.

Opening and closing the door without using the external compensation coils (i.e., exposing the inner shield layer to the ambient field) causes distortions of few nT in the center of the MSR.  
This residual field is easily removed by degaussing only the inner shield.
The response to external field variations was also studied by applying a sequence of external-field pulses of several second duration with short rise time and $\sim 100$ $\mu$T amplitude. After five pulses, the residual field in the center of the room returned to its initial value to within the measurement precision of the fluxgate magnetometer.

The residual field in the center of the MSR is a superposition of fields emerging from the inner shield layer and contributions of `imperfections' due to overlaps of $\mu$-metal sheets, field penetrations of the feedthroughs on the wall, and magnetic parts inside the MSR.
The dependence of the magnetic field magnitude as function of the distance from the wall is shown in Fig.~\ref{fig:fielddistance} for feedthroughs of various sizes. 
After repeated degaussing of the inner shield, the residual-field-distribution is reproducible at the fluxgate sensor resolution.

For feedthroughs with diameter $<$~60~mm, the field decreases rapidly with distance and is $<$~3~nT at a distance of 50~mm from the inner $\mu$-metal wall. 
As expected, the field drops more slowly for the 130 mm diameter feedthroughs.
This curve shows a bump at 15 cm distance due to the presence of a fixed magnetic object nearby.

Figure~\ref{fig:fielddistance2}(a) shows the horizontal  X-Y projection of the field  in a plane that extends outward from the closed door at the vertical center of the chamber with the overlap region as indicated by the arrows.  
Whereas the overlap of the $\mu$-metal wall effectively reduces the magnetic resistance between the shield wall and the door, the door cannot be degaussed as good as the main material of the shield.  
Fig.~\ref{fig:fielddistance2} (a) shows that the magnetic field penetrates into the room at the door-wall overlap. The $\mu$-metal side wall at X = 0 attracts the field and keeps it away from the center, therefore at 0.6~m away from the door wall is below 1~nT, see Fig.~\ref{fig:fielddistance2} (b).  
For the residual fields, the contact of the $\mu$-metal of the door and the $\mu$-metal of the wall is limiting the residual fields inside the room.
It is a difficult mechanical problem to get sufficient contact between the $\mu$-metal layers without air gaps.
Even a very small localized airgap is critical for the residual field.

\subsection{\label{precession} Spin-precession measurements}

\begin{figure}
\includegraphics[width=1\linewidth]{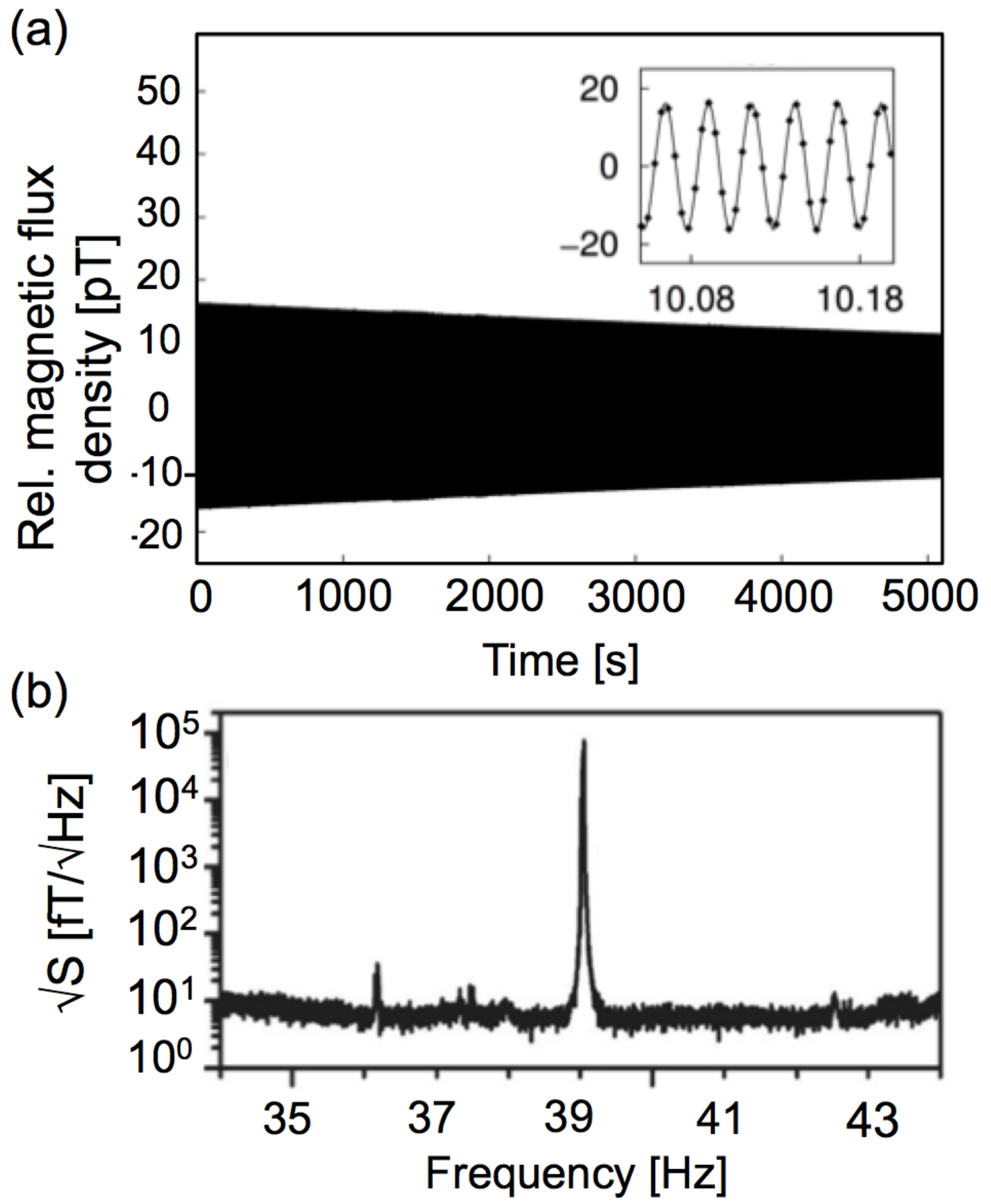}
\caption{\label{fig:signal} (a) Measured SQUID signal of the precessing $^3$He nuclear spins, filtered by a 7th order Butterworth band pass filter (35 to 45 Hz). The inset in the figure is a zoom into the data set. (b) Magnetic flux density spectrum of the measured precession signal, unfiltered. The 39~Hz peak is the Larmor frequency of $^3$He in a 1.2 $\mu$T guiding field. The Xe peak is outside the shown freqency band. The small peak at 36.2~Hz could be caused by a ventilation fan.
}
\end{figure}

The homogeneity of the magnetic field will strongly influence the coherence time of the nuclear spin during a magnetic resonance measurement.
Therefore, the transverse relaxation time $T_2^*$ can be used to probe the field homogeneity.
Figure~\ref{fig:signal}a shows the Larmor precession of hyperpolarized xenon nuclei in a applied magnetic field of 1.2 $\mu$T inside the MSR. 

The $^3$He gas was in a 3 cm, nominally spherical sealed glass bulb with 500~mbar $^3$He, 100~mbar of xenon (90\% $^{129}$Xe), and 50~mbar of N$_2$ introduced as a buffer gas. The noble gases were hyperpolarized by spin exchange optical pumping~\cite{SEOP} and manufactured at the University of Michigan.
The distance between the center of the cell and the LTc SQUID magnetometer 31~\cite{drung2003} during the measurement was 11~cm. 
The white noise level of the system is $<10$ fT/$\sqrt{Hz}$. 
The sampling rate was 250 Hz. 
The MSR was degaussed and then the bulb with the polarized gas was passed by hand through  a 12~cm hole in the MSR rear wall and set at the center of the guide field.
The free-precession was initiated by a 90$^\circ$ non-adiabatic change of the guiding field direction ($\sim$ 10 ms switching time).

A transverse relaxation time $T_2^*$ up to 16900~s was observed. 
Contributions to dephasing from atom-atom collisions and atom-wall collisions, which are usually less significant than the field gradient effect, now become important inside the MSR.

In addition to the relaxation time, the signal-to-noise ratio of the precession signal plays a pivotal role in determining the precision of the frequency by limiting the observation time. 
The Fourier transform of the precession signal is shown in Figure~\ref{fig:signal}(b).

\section{\label{conclusion}Summary}
%
We have described the design principles, construction and performance of a portable magnetically-shield room. The residual static field of the two-$\mu$-metal layer and one RF-layer room is $<1$~nT and has a field gradient of less than 300~pT/m. 
This exceeds the performance of comparable non-transportable MSRs significantly. 
The shielding characteristics, initially measured at the factory in Switzerland, were reproduced after transporting the assembled room from the factory to a laboratory in Garching, Germany and then again after another move to the FRM-II reactor site about 1 km away. 
 We have demonstrated the performance of the MSR by SQUID-based measurements of the precession of the nuclear magnetic moment of $^3$He gas in a glass bulb of 30~mm in diameter in an applied field of 1.2 $\mu$T. Although the transverse relaxation time $T_2^*$ is reduced by a number of factors in addition to the field homogeneity, we achieved relaxation times of $T_2^*$=16900~s. 
This outstanding performance of the experimental environment inside the MSR will serve as the basis of forthcoming experiments aiming at, e.g., the determination of the electric dipole moment of the neutron (nEDM) and noble-gas nuclei.  
Field homogeneity over scales of a meter meets the performance needed for the nEDM experiment. 
For the spin-precession experiments, the field characteristics over a much smaller distance are relevant. 
This shielded room opens up new opportunities in a variety of fields from precision measurements in physics to biomedical applications to materials studies.

We acknowledge the support at the FRM-II, the workshop of the Max-Planck-Insititute for Plasma-Physics and at the TUM Physics Department in Garching. 
This work was supported by the DFG Priority Program SPP 1491 and the DFG Cluster of Excellence 'Origin and Structure of the Universe', as well as by, in part, the U.S. National Science Foundation under grant PHY1205671 and DOE grant DE FG02 04 ER41331.

\end{document}